\documentclass[10pt,final,journal,a4paper,oneside,twocolumn]{IEEEtran}

\usepackage{subfigure}
\usepackage{srcltx}
\usepackage{psfrag}
\usepackage{epsfig}
\usepackage{color}
\usepackage[tbtags,sumlimits,nointlimits,reqno]{amsmath}
\usepackage{amssymb}
\usepackage{showkeys}   
\usepackage{color}
\usepackage{float}
\usepackage{mathtools, cuted}
\usepackage[makeroom]{cancel}
\usepackage[normalem]{ulem}

\begin{document}

\title{Floquet-Mode Solutions of \\Space-Time Modulated Huygens' Metasurfaces}

\author{%
       Shulabh~Gupta, Tom. J. Smy and Scott A. Stewart 
\thanks{S.~Gupta, T. J. Smy and  S. A.~Stewart are with the Department of Electronics, Carleton University, Ottawa, Ontario, Canada. Email: shulabh.gupta@carleton.ca}
}
\markboth{MANUSCRIPT DRAFT}
{Shell \MakeLowercase{\textit{et al.}}: Bare Demo of IEEEtran.cls for Journals}

\maketitle
\begin{abstract}
A rigorous Floquet mode analysis is proposed for a zero thickness space-time modulated Huygens' metasurface to model and determine the strengths of the new harmonic components of the scattered fields. The proposed method is based on Generalized Sheet Transition Conditions (GSTCs) treating a metasurface as a spatial discontinuity. The metasurface is described in terms of Lorentzian electric and magnetic surface susceptibilities, $\chi_\text{ee}$ and $\chi_\text{mm}$, respectively, and its resonant frequencies are periodically modulated in both space and time. The unknown scattered fields are then expressed in terms of Floquet modes, which when used with the GSTCs, lead to a system of field matrix equations. The resulting set of linear equations are then solved numerically to determine the total scattered fields. Using a finite-difference time domain (FDTD) solver, the proposed method is validated and confirmed for several examples of modulation depths ($\Delta_p$) and frequencies ($\omega_p$). Finally, the computed steady-state scattered fields are Fourier propagated analytically, for visualization of refracted harmonics. The proposed method is simple and versatile and able to determine the steady-state response of a space-time modulated Huygen's metasurface, for arbitrary modulation frequencies and depths.
\end{abstract}

\begin{keywords}
Electromagnetic Metasurfaces, Electromagnetic Propagation, Floquet Analysis, Explicit Finite-Difference, Generalized Sheet Transition Conditions (GSTCs), Lorentz Dispersions, Parametric Systems.
\end{keywords}


\section{Introduction}

Recently, there has been a strong growing interest in Huygen's metasurfaces due to their impedance matching capabilities with free-space and their versatile applications in wavefront shaping \cite{Elliptical_DMS}\cite{GeneralizedRefraction}\cite{meta3}. They are constructed using a 2-D array of electrically small Huygen's sources, exhibiting perfect cancellation of backscattered fields, due to optimal interactions of their electric and magnetic dipolar moments \cite{Kerker_Scattering}. Some efficient implementations of Huygens' metasurfaces are based on all-dielectric resonators \cite{Kivshar_Alldielectric}\cite{Elliptical_DMS}\cite{AllDieelctricMTMS} and orthogonally collocated small electric and magnetic dipoles \cite{Grbic_Metasurfaces}\cite{HuygenBook_Eleftheriades}.

While the majority of the work on metasurfaces has been focussed on static (linear time invariant) metasurfaces, there is also a growing interest in dynamic metasurfaces, where the constitutive parameters of the metasurface unit cells are controlled in real-time. An important class of such dynamic electromagnetic structures is \emph{space-time modulated metasurfaces}, where their constitutive parameters are periodically modulated in both space and time, at comparable frequency scales to the input excitations. This leads to a complex interaction of the incident wavefronts with the metasurfaces resulting in exotic effects such as generation of new harmonic components and Lorentz non-reciprocity \cite{STGradMetasurface}\cite{ShaltoutSTMetasurface}. Space-time modulated metasurfaces fall within the general framework of space-time modulated mediums \cite{TamirST}\cite{OlinerST}, which have found important applications in acousto-optical systems for spectrum analysis, parametric oscillators and amplifiers, for instance \cite{Goodman_Fourier_Optics}\cite{Saleh_Teich_FP}\cite{TamirAcoustoDiffraction}. 

Consequently, a combination of the wave-shaping capabilities of Huygens' metasurfaces with space-time modulation principles, is an interesting avenue to explore for advanced electromagnetic wave control, in both space and time. To investigate into the properties of space-time modulated Huygens' metasurfaces, a finite-difference time-domain (FDTD) technique has recently been proposed to analyze a zero thickness model of Huygens' metasurfaces \cite{Smy_Metasurface_Linear}\cite{Stewart_Metasurface_STM}, based on Generalized Sheet Transition Conditions (GSTCs) \cite{KuesterGSTC}. In contrast to the other existing techniques for analyzing static metasurfaces in frequency domain \cite{CalozFDTD}\cite{Caloz_MetaModelling}, the FDTD analysis is naturally applicable to the problem of space-time modulated metasurfaces, considered herein. 

While such numerical methods are useful to determine the time-evolution of scattered waves from space-time modulated metasurfaces for a given input wave, efficient determination of steady-state response of the metasurface is an equally important problem to solve. This issue is addressed in this work, whereby exploiting the periodic nature of the spatio-temporal perturbation on the metasurface, the scattered fields are expressed in terms of Floquet modes. The Huygens' metasurface is modelled using surface susceptibilities following a physically motivated Lorentzian profile, whose resonant frequencies are parametrized to emulate a space-time modulation of the metasurface. Combined with GSTCs, the Floquet mode amplitudes are computed by solving a set of linear equations. The proposed method thus efficiently computes the steady-state response of a zero-thickness space-time modulated Huygens' metasurface, consistent with the FDTD field solutions. Furthermore, integrating the proposed method with analytical Fourier methods \cite{Goodman_Fourier_Optics}, the propagation of scattered fields are conveniently visualized in free-space.

The paper is structured as follows. Section II describes the problem statement of this work, and develops the fields equations governing the scattering fields from a space-time modulated metasurface based on GSTCs and Lorentz surface susceptibilities. Section III presents the proposed method based on Floquet mode expansions, forming the set of linear equations to be solved numerically. Several results are then presented for both time-only and space-time modulated metasurfaces. Conclusions are provided in Sec. IV, and a brief summary of the FDTD method is provided in Appendix V, for the sake of self-consistency and completeness of the paper.

\section{Space-Time Modulated Metasurfaces}

\subsection{Problem Statement}

\begin{figure}[htbp]
\begin{center}
\psfrag{a}[c][c][0.8]{$z=0$}
\psfrag{z}[c][c][0.8]{$z$}
\psfrag{x}[c][c][0.8]{$x$}
\psfrag{t}[c][c][0.8]{$t$}
\psfrag{b}[c][c][0.8]{$\psi(\mathbf{r}, t) = \psi_0(\mathbf{r}, t) \sin(\omega_0t)$}
\psfrag{f}[c][c][0.6]{$\boxed{\chi_{ee}(x, t),\; \chi_{mm}(x, t)}$}
\psfrag{d}[c][c][0.6]{$\psi_2(\mathbf{r}, t) \sin\{(\omega_0 - \omega_p)t\}$}
\psfrag{c}[l][c][0.6]{$\psi_1(\mathbf{r}, t) \sin\{\omega_0t\}$}
\psfrag{e}[c][c][0.6]{$\psi_3(\mathbf{r}, t) \sin\{(\omega_0 + \omega_p)t\}$}
\includegraphics[width=0.75\columnwidth]{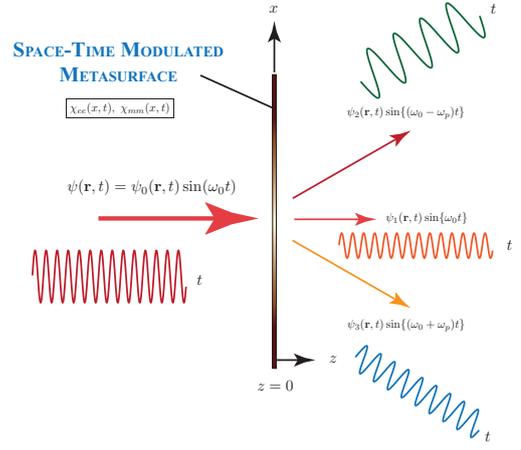}
\caption{A general Illustration of a space-time modulated Huygens' metasurface under normally incident CW plane-wave resulting in generation of several frequency harmonics, refracted along different angles. $\omega_p$ is the pumping frequency.}\label{Fig:Problem}
\end{center}
\end{figure}

Metasurfaces are zero thickness electromagnetic structures that act as discontinuities in space. The exact zero thickness nature of electromagnetic structures was developed by Idemen in terms of Generalized Sheet Transition Conditions (GSTCs) \cite{IdemenDiscont}, which were later applied to metasurfaces \cite{KuesterGSTC}. Consider a Huygens' metasurface illustrated in Fig.~\ref{Fig:Problem}, where the metasurface lies in the $x-y$ plane at $z=0$, with a wave incidence on the left, normal to the surface. The wave interacts with the metasurface and produces a transmitted and a reflected wave, along the forward and backward direction, respectively. This interaction of the metasurface with the electromagnetic waves is described by the GSTCs, using the transverse components of electric and magnetic surface polarizabilities $\mathbf{P}$ and $\mathbf{M}$, as \cite{Metasurface_Synthesis_Caloz}

\begin{subequations}\label{Eq:GSTC}
\begin{equation}
\hat{\mathbf{z}}\times\Delta \mathbf{H}(x,t) = \frac{d\mathbf{P}_{||}(x,t)}{dt}
\end{equation}
\begin{equation}
\Delta \mathbf{E}(x,t)\times \hat{\mathbf{z}} = \mu_0\frac{d\mathbf{M}_{||}(x,t)}{dt},
\end{equation}
\end{subequations}

\noindent where 

\begin{align}
\Delta \mathbf{E} =  (\mathbf{E}_t -  \mathbf{E}_0 -  \mathbf{E}_r),\quad
\Delta \mathbf{H} =  (\mathbf{H}_t -  \mathbf{H}_0 -  \mathbf{H}_r). \notag
\end{align}

\noindent The surface polarizabilities on the Huygens' metasurface are related to the average fields around the metasurface and can be described in terms of scalar electric and magnetic surface susceptibilities $\chi_\text{ee}$ and $\chi_\text{mm}$, respectively, as\footnote{Assuming no rotation of polarization and only one component of the fields along the principal axis.}.

\begin{subequations}
\begin{equation}
\mathbf{\tilde{Q}}_{||}(\omega) =  \tilde{\chi}_\text{ee} \mathbf{\tilde{E}}_\text{av}(\omega),
\end{equation}
\begin{equation}
\mathbf{\tilde{M}}_{||}(\omega) =  \tilde{\chi}_\text{mm} \mathbf{\tilde{H}}_\text{av}(\omega), 
\end{equation}
\end{subequations}

\begin{equation}
\text{where}~\mathbf{\tilde{E}}_\text{av} = \left[ \frac{\mathbf{\tilde{E}}_0 + \mathbf{\tilde{E}}_t + \mathbf{\tilde{E}}_r}{2}\right],~\mathbf{\tilde{H}}_\text{av} = \left[ \frac{\mathbf{\tilde{H}}_0 + \mathbf{\tilde{H}}_t + \mathbf{\tilde{H}}_r}{2}\right]\notag
\end{equation}

\noindent and $\mathbf{Q} = \mathbf{P}/\epsilon_0$ is the normalized electric polarizability\footnote{The fields expressions with a tilde, $\tilde{\psi}$ denote the frequency domain quantities.}.

Next, consider a space-time modulated metasurface, whose electric and magnetic susceptibilities are both a function of space and time, i.e. $ \chi_\text{ee}(x, t)$ and $ \chi_\text{mm}(x, t)$. Let us restrict here to a periodic modulation only, with a pumping frequency $\omega_p$ and the spatial frequency $\beta_p$. Due to this periodic spatio-temporal perturbation on the metasurface, assumed to be infinite in size, the scattered transmission (and reflection) fields can be expressed in terms of Floquet series as

\begin{equation}\label{Eq:TotalTFields}
E_t(x, z= 0_+, t) = \sum_{n=0}^\infty q_n e^{j\omega_n t} e^{j\beta_n x},
\end{equation}

\noindent where $\omega_n = \omega_0 + n\omega_p$, and $\beta_n = \beta_{x0} + n\beta_p$ with $\beta_{x0}=0$ due to assumed normal input incidence. Each harmonic term of this expansion with a temporal frequency $\omega_n$, represents an oblique forward propagating plane-wave in the $+z$ direction, and making an angle $\theta_n$ measured from the normal of the metasurface, as illustrated in Fig.~\ref{Fig:Problem}. These refraction angles are given by

\begin{equation}\label{Eq:BetapRefract}
\theta(\omega_n) = \sin^{-1}\left[\frac{n}{(1 + n\omega_p/\omega_0) }\frac{\beta_p}{k_0}\right],
\end{equation}

\noindent where $k_0$ is the free-space wavenumber of the fundamental frequency. Based on this simple physical argument, the angle of refraction of these newly generated frequency components from a space-time modulated metasurface, can be determined. However, the strengths of these harmonics (weights $q_n$) are still unknown and must be determined taking the exact electromagnetic interaction of the waves with the metasurface into account using the GSTCs of \eqref{Eq:GSTC}.

\subsection{Field Equations}

The space-time modulated problem considered here is naturally treated in the time domain. Consequently, the surface susceptibilities must be defined in time domain as well. The most common, and causal, description of the susceptibilities is in terms of Lorentz dispersion, which is also typical of the Huygens' sources used to construct the metasurfaces \cite{Kivshar_Alldielectric}\cite{PDM_Gupta}\cite{Huygen_Microwave_Achouri}. Consequently, the electric and magnetic susceptiblities can be expressed as 

\begin{subequations}\label{Eq:Lorentz}
\begin{equation}
\tilde{\chi}_\text{ee}(\omega)  =  \frac{\omega_{ep}^2}{(\omega_{e0}^2 - \omega^2) + i\alpha_e \omega} 
\end{equation}
\begin{equation}
\tilde{\chi}_\text{mm}(\omega)  =  \frac{\omega_{mp}^2}{(\omega_{m0}^2 - \omega^2) + i\alpha_m \omega},  
\end{equation}
\end{subequations}

\noindent where $(\omega_{e0}$, $\omega_{m0}),\; (\omega_{ep}$, $\omega_{mp}),\; (\alpha_e, \alpha_m)$ are the electric and magnetic resonant frequencies, plasma frequencies and loss coefficients, respectively.

The space-time modulation can now be introduced into the metasurface, by sinusoidally modulating, for instance, the resonant frequencies of the two Lorentzian susceptibilities, as \cite{Stewart_Metasurface_STM}

\begin{subequations}\label{Eq:w0Modulation}
\begin{equation}
\omega_{e0}(x, t) = \omega_{e0} \{1 + \Delta_e\cos(\omega_p t + \beta_px)\}
\end{equation}
\begin{equation}
\omega_{m0}(x, t) = \omega_{m0} \{1 + \Delta_m\cos(\omega_pt + \beta_px)\},
\end{equation}
\end{subequations}

\noindent where $\Delta_e$ and $\Delta_m$ are the modulation depths and the spatial frequencies of the perturbation, for electric and magnetic resonances, respectively. They can now be inserted into the GSTCs of \eqref{Eq:GSTC}, for a prescribed input fields, to determine the total scattered fields.

To illustrate the procedure, let us assume a normally incident plane-wave ($|\mathbf{E_0}| =\text{const.}$) where the corresponding transmitted and reflected fields, given by

\begin{align}\label{Eq:FieldConvention}
&\mathbf{E}_0(z,t) = E_0e^{j(\omega_0t - k_0z)}~\mathbf{\hat{y}},~ \mathbf{H}_0(z,t) = \frac{\mathbf{\hat{z}}  \times\mathbf{E}_0(z,t)}{\eta_0} \\
&\mathbf{E}_t(z,t) = E_t(x,t)e^{-j k_0z}~\mathbf{\hat{y}},~ \mathbf{H}_t(z,t) = \frac{\mathbf{\hat{z}}  \times\mathbf{E}_t(z,t)}{\eta_0} \notag\\
&\mathbf{E}_r(z,t) = E_r(x,t)e^{+j k_0z}~\mathbf{\hat{y}},~ \mathbf{H}_r(z,t) = \frac{\mathbf{E}_r(z,t)\times \mathbf{\hat{z}}  }{\eta_0}.\notag
\end{align}

\noindent Each of these field components are related to their respective polarizabilities on the metasurface (at $z=0$) through the Lorentz susceptibilities of \eqref{Eq:Lorentz} with \eqref{Eq:w0Modulation}, which can then be expressed in the time domain as\footnote{for the lossless case, for simplicity.}  \cite{Smy_Metasurface_Linear}

\begin{subequations}\label{Eq:TD_Lorentz}
\begin{equation}
\frac{d^2Q_i}{dt^2} +\omega_{e0}^2\{1 + \Delta_e\cos(\omega_p t + \beta_px)\}^2Q_i = \omega_{ep}^2E_i(t)
\end{equation}
\begin{equation}
\frac{d^2M_i}{dt^2} +\omega_{m0}^2\{1 + \Delta_m\cos(\omega_p t + \beta_px)\}^2M_i = \frac{\omega_{ep}^2}{\eta_0}E_i(t),\footnote{with proper sign of the right hand side term, depending on incident ($-$), reflected ($+$) or transmitted fields ($-$), according to \eqref{Eq:FieldConvention}.}
\end{equation}
\end{subequations}

\noindent where the subscript $i = 0, t, r$ for incident, transmitted and reflected fields, respectively. Finally, all the scattered fields are related to their polarizabilities following the GSTCs of \eqref{Eq:GSTC}, as

\begin{subequations}\label{Eq:GSTCEquations}
\begin{equation}
\frac{dQ_0}{dt} + \frac{dQ_t}{dt} + \frac{dQ_r}{dt} = \frac{2}{\eta_0\epsilon_0}(E_0 - E_r - E_t),
\end{equation}
\begin{equation}
\frac{dM_0}{dt} + \frac{dM_t}{dt} + \frac{dM_r}{dt} = \frac{2}{\mu_0}(E_t - E_r - E_0) 
\end{equation}
\end{subequations}

Equations \eqref{Eq:TD_Lorentz} and \eqref{Eq:GSTCEquations} thus represent two sets of field equations, that must be solved to determine the transmitted and reflected fields, $E_t(x,z, t)$ and $E_r(x,z, t)$, in steady state, for a given input excitation $\mathbf{E_0} = E_0e^{j\omega_0t}~\mathbf{\hat{y}}$ at the input of the metasurface at $z=0_-$.

\begin{figure*}
\begin{align}\label{Eq:FieldEquations}
&\sum_n \left[\Delta_{e1}a_{n-2} +  \Delta_{e2} a_{n-1} +  A_{e, n} a_n + \Delta_{e2} a_{n+1} + \Delta_{e1}a_{n+2} \right]e^{jn\Omega t}  = E_0~\quad \text{(Incident E-Fields)} \\
&\sum_n \left[\Delta_{e1}b_{n-2} +  \Delta_{e2} b_{n-1} +  A_{e, n}b_n + \Delta_{e2} b_{n+1} + \Delta_{e1}b_{n+2} -q_n \right]e^{jn\Omega t}  = 0~\quad \text{(Transmitted E-Fields)} \notag\\
&\sum_n \left[\Delta_{e1}c_{n-2} + \Delta_{e2} c_{n-1} +  A_{e, n}c_n + \Delta_{e2} c_{n+1} + \Delta_{e1}c_{n+2} -p_n \right]e^{jn\Omega t}  = 0~\quad \text{(Reflected E-Fields)} \notag\\  
&\sum_n \left[\Delta_{m1}d_{n-2} + \Delta_{m2} d_{n-1} +  A_{m,n}d_n + \Delta_{m2} d_{n+1} + \Delta_{m1}d_{n+2} \right]e^{jn\Omega t}  = -\frac{E_0}{\eta_0}~\quad \text{(Incident H-Fields)} \notag\\
&\sum_n \left[\Delta_{m1}e_{n-2} + \Delta_{m2} e_{n-1} +  A_{m,n}e_n + \Delta_{m2} e_{n+1} + \Delta_{m1} e_{n+2} -b_n \right]e^{jn\Omega t}  = 0~\quad \text{(Transmitted H-Fields)} \notag\\
&\sum_n \left[\Delta_{m1}f_{n-2} + \Delta_{m2} f_{n-1} +  A_{m,n}f_n + \Delta_{m2} f_{n+1} + \Delta_{m1} f_{n+2} -f_n \right]e^{jn\Omega t}  = 0~\quad \text{(Reflected H-Fields)} \notag\\  
&\sum_n \left[ 2cB_{n} a_n + 2cB_{n} b_n + 2cB_{n} c_n + p_n + q_n \right]e^{jn\Omega t} = E_0~\quad \text{(GSTC Equation)}\notag\\
&\sum_n \left[ \frac{\mu_0B_{n}}{2} d_n + \frac{\mu_0B_{n}}{2} e_n + \frac{\mu_0B_{n}}{2} f_n + p_n - q_n \right]e^{jn\Omega t} = -E_0~\quad \text{(GSTC Equation)}\notag
 \end{align}\rule[1ex]{18cm}{0.5pt} \\
 \begin{align}\label{Eq:FieldMatrix}
&\overbrace{\left[ \begin{array}{cccc cccc}
\mathbf{S_e} & 0 & 0 & 0 & 0 & 0 & 0 & 0  \\
0 & \mathbf{S_e} & 0 & 0 & 0& 0  & 0 & -\mathbf{I}  \\
0 & 0 & \mathbf{S_e} & 0 & 0 & 0 & -\mathbf{I} & 0  \\
0 & 0 & 0 & \mathbf{S_m} & 0 & 0 & 0 & 0  \\
0 & 0 & 0 & 0 & \mathbf{S_m} & 0 & 0  & \mathbf{I}/\eta_0  \\
0 & 0 & 0 & 0& 0 & \mathbf{S_m} & -\mathbf{I}/\eta_0  & 0 \\
\mathbf{S_g}/(2c_0) & \mathbf{S_g}/(2c_0) & \mathbf{S_g}/(2c_0) & 0& 0  & 0 & \mathbf{I} & \mathbf{I}\\
0 & 0 & 0 & \mathbf{S_g} & \mathbf{S_g} & \mathbf{S_g} & \mathbf{I}  & -\mathbf{I}
 \end{array} \right]}^{\mathbf{C_n}}
\overbrace{\left[ \begin{array}{c}
\mathbf{a_n} \\
\mathbf{b_n}\\
\mathbf{c_n}\\
\mathbf{d_n}\\
\mathbf{e_n}\\
\mathbf{f_n}\\
\mathbf{p_n}\\
\mathbf{q_n}
 \end{array} \right]}^{\mathbf{V_n}} 
 =
\overbrace{\left[ \begin{array}{c}
\mathbf{I_0} \\
0\\
0\\
-\mathbf{I_0}/\eta_0\\
0\\
0\\
\mathbf{I_0} \\
-\mathbf{I_0} 
 \end{array} \right] , \text{where}~\quad \mathbf{I_0} = \left[ \begin{array}{c}
\vdots\\
0\\
0\\
1\\
0\\
0\\
\vdots 
 \end{array} \right]}^{\mathbf{E_n}} \notag\\
&\mathbf{S_e} = \left[ \begin{array}{cccccc} 
A_{e, -N} & \Delta_{e2} & \Delta_{e1} & 0 & \cdots & 0   \\
\vdots & \vdots & \vdots& \vdots & \vdots  & \cdots  \\
\Delta_{e2}  & A_{e, -1}  & \Delta_2 & \Delta_{e1} & \cdots & 0   \\
\Delta_{e1}  & \Delta_{e2} & A_{e,0}  & \cdots & \Delta_{e2} & \Delta_{e1}   \\
0 & \cdots & \Delta_{e1} & \Delta_{e2} & A_{e, +1}  & \Delta_{e2} \\
& \cdots \vdots & \vdots & \vdots& \vdots & \vdots   \\
0 & \cdots  & 0 & \Delta_{e1} & \Delta_{e2} & A_{e, +N}  
\end{array} \right]
, \quad
\mathbf{S_m} = \left[ \begin{array}{cccccc} 
A_{m, -N} & \Delta_{m2} & \Delta_{m1} & 0 & \cdots & 0   \\
\vdots & \vdots & \vdots& \vdots & \vdots  & \cdots  \\
\Delta_{m2}  & A_{m, -1}  & \Delta_2 & \Delta_{m1} & \cdots & 0   \\
\Delta_{m1}  & \Delta_{m2} & A_{m, 0}  & \cdots & \Delta_{m2} & \Delta_{m1}   \\
0 & \cdots & \Delta_{m1} & \Delta_{m2} & A_{m, +1}  & \Delta_{m2} \\
& \cdots \vdots & \vdots & \vdots& \vdots & \vdots   \\
0 & \cdots  & 0 & \Delta_{m1} & \Delta_{m2} & A_{m, +N}  
\end{array} \right]\notag\\
&\mathbf{S_g} = \text{diag}\{B_{-N}, B_{-N+1},\cdots, B_{-1}, B_0, B_1, \cdots, B_{N-1}, B_N\}.
\end{align}\rule[1ex]{18cm}{0.5pt}
\end{figure*}

\section{Proposed Floquet Mode Expansion Solutions}

\subsection{Field Matrix Equations}

Let us consider a metasurface with a periodic space-time modulation of the resonant frequencies following \eqref{Eq:w0Modulation}, and expand the unknown reflected and transmitted fields using Floquet expansion as

\begin{subequations}\label{Eq:EtEt_Expansion}
\begin{equation}
E_t(t) =  \sum_n q_ne^{j\omega_0 t}e^{jn(\omega_pt + \beta_px)}  = \sum_n q_ne^{j\omega_0 t}e^{jn\Omega},
\end{equation}
\begin{equation}
E_r(t) = \sum_n p_ne^{j\omega_0 t}e^{jn(\omega_pt + \beta_px)} = \sum_n p_ne^{j\omega_0 t}e^{jn\Omega},
\end{equation}
\end{subequations}

\noindent where a new variable $\Omega = (\omega_p t + \beta_p x)$ is introduced for compact notation. Similarly, the corresponding unknown polarizabilities, can also be expanded in Floquet series as

\begin{align}\label{Eq:QM_Expansion}
&Q_0(t) = \sum_n a_ne^{j\omega_0 t}e^{jn\Omega}, ~\quad M_0(t) = \sum_n d_ne^{j\omega_0 t}e^{jn\Omega}, \notag\\
&Q_t(t) = \sum_n b_ne^{j\omega_0 t}e^{jn\Omega}, ~\quad M_t(t) = \sum_n e_ne^{j\omega_0 t}e^{jn\Omega}, \notag\\
&Q_r(t) = \sum_n c_ne^{j\omega_0 t}e^{jn\Omega}, ~\quad M_r(t) = \sum_n f_ne^{j\omega_0 t}e^{jn\Omega}.\notag
\end{align}

\noindent Next, substituting the above expressions in \eqref{Eq:TD_Lorentz} and \eqref{Eq:GSTCEquations} and re-arranging the terms, for incident, transmitted and reflected E- and H-fields, we get \eqref{Eq:FieldEquations}, where

\begin{align}
&\Delta_{e1}= \frac{\Delta_e^2\omega_{e0}^2}{4\omega_{ep}^2},\quad \Delta_{e2}= \frac{\Delta_e\omega_{e0}^2}{\omega_{ep}^2}, \notag\\
& A_{e, n} = \left\{\frac{\omega_{e0}^2}{\omega_{ep}^2} - \frac{(\omega_0 + n\omega_p)^2}{\omega_{ep}^2} +  \frac{\Delta_e^2\omega_{e0}^2}{2\omega_{ep}^2}\right\}\notag.\\
&\Delta_{m1}= \frac{\Delta_m^2\omega_{m0}^2}{4\omega_{mp}^2},\quad \Delta_{m2}= \frac{\Delta_m\omega_{m0}^2}{\omega_{mp}^2}, \notag\\
& A_{m,n} = \left\{\frac{\omega_{m0}^2}{\omega_{mp}^2} - \frac{(\omega_0 + n\omega_p)^2}{\omega_{mp}^2} +  \frac{\Delta_m^2\omega_{m0}^2}{2\omega_{mp}^2}\right\}\notag\\
& B_{n} =  j(\omega_0 + jn\omega_p).
\end{align}

\noindent These equations are obtained by expressing the cosine function using Euler's form and then grouping the terms of common complex exponentials. Each of these series equations, can be truncated to $2N+1$ harmonic terms\footnote{assuming that the harmonic amplitudes rapidly falls to zero with increasing index $n$, which is the case in all the forthcoming results.}, and be written in a matrix form as shown in \eqref{Eq:FieldMatrix}. In compact form, the matrix equation \eqref{Eq:FieldMatrix} can be written as

\begin{equation}
[\mathbf{C_n}][\mathbf{V_n}] = [\mathbf{E_n}],
\end{equation}

\noindent which represents $8\times (2N+1)$ linear equations with same number of unknowns. The sought field solutions can now be numerically computed using this matrix equation as

\begin{equation}
[\mathbf{V_n}] = \text{inv}\{[\mathbf{C_n}]\} [\mathbf{E_n}],
\end{equation}

\noindent from which the corresponding $p_n$'s and $q_n$'s can be extracted to construct the reflected and transmitted fields, respectively, following \eqref{Eq:EtEt_Expansion}.

\subsection{Results}

To validate the results of the proposed method, an FDTD solver is used, which has been recently proposed in \cite{Smy_Metasurface_Linear}\cite{Stewart_Metasurface_STM} to directly simulate \eqref{Eq:TD_Lorentz} and \eqref{Eq:GSTCEquations}, for an arbitrary time-domain input. A brief summary of the method is described in the appendix for self-consistency of the paper. In the results described next, the modulation depths of the electric and magnetic resonant frequencies are assumed to be equal for simplicity, i.e. $\Delta_e=\Delta_m= \Delta_p$. Furthermore, the output fields from the proposed method are constructed using a Gaussian envelopes\footnote{This provides a finite bandwidth in the temporal FFTs as opposed to ideal delta functions which are difficult to capture in numerical computations. These Gaussian envelopes should not be seen as the input pulse shapes, but rather as windowing function.}, for better conditioning of the numerical Fourier transforms and for a closer comparison with the FDTD method, given by

\begin{align}\label{Eq:FieldConstruction}
E_t(x,t) = \sum_{n= -N}^{+N} q_n \exp\left\{ - \left(\frac{t}{T_0}\right)^2\right\} \cos(\omega_n t + \beta_n x),\notag\\
E_r(x,t) = \sum_{n = -N}^{+N} p_n \exp\left\{ - \left(\frac{t}{T_0}\right)^2\right\} \cos(\omega_n t + \beta_n x).
\end{align}

\begin{figure}[htbp]
\begin{center}
\psfrag{C}[c][c][0.8]{frequency, $\omega/\omega_0$}
\psfrag{A}[c][c][0.8]{$\mathcal{F}_t\{E_t(t)\}$}
\psfrag{B}[c][c][0.8]{$\mathcal{F}_t\{E_r(t)\}$}
\psfrag{E}[l][c][0.6]{FDTD}
\psfrag{F}[l][c][0.6]{Floquet}
\psfrag{X}[c][c][0.8]{$\Delta_p = 0.05$, $f_m = 0.05f_0$}
\psfrag{Y}[c][c][0.8]{$\Delta_p = 0.1$, $f_m = 0.1f_0$}
\psfrag{Z}[c][c][0.8]{$\Delta_p = 0.25$, $f_m = 0.25f_0$}
\includegraphics[width=\columnwidth]{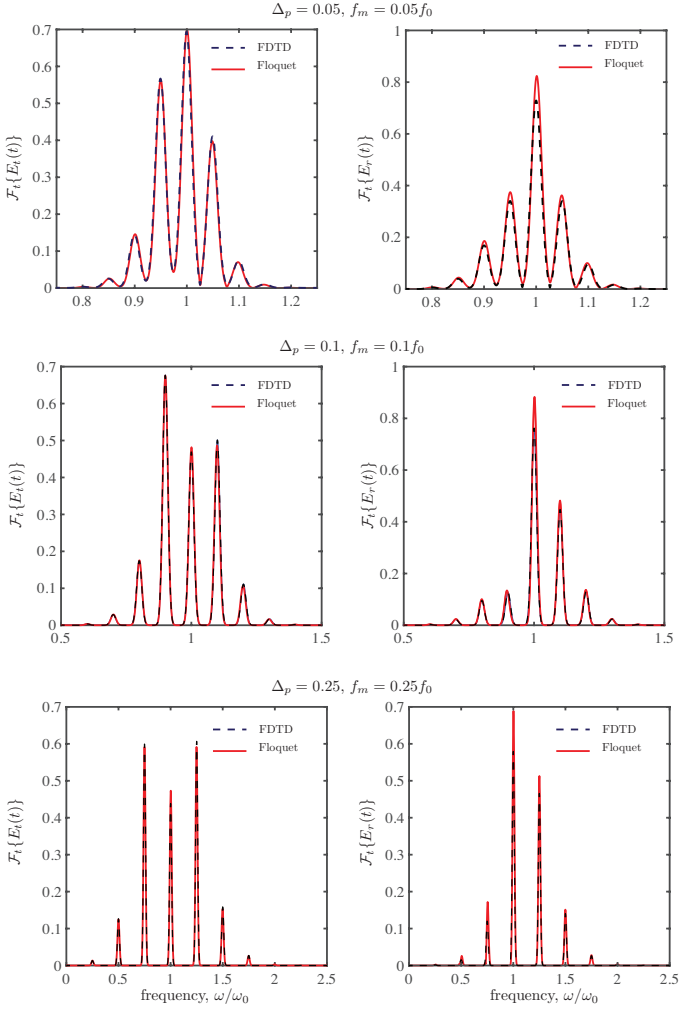}
\caption{Spectrum of transmission and reflection fields from an infinite time-only ($\beta_p=0$) modulated metasurface incident with a normally incident plane-wave, for the case of a) $\Delta_p = 0.05$, $\omega_p = 0.05\omega_0$, b) $\Delta_p = 0.1$, $\omega_p = 0.1\omega_0$ and c) $\Delta_p = 0.25$, $\omega_p = 0.25\omega_0$. Total number of harmonics $2N+1 = 61$ and $T_0 =100~$fs in \eqref{Eq:FieldConstruction}. The metasurface parameters are: $\omega_{e0} = 2\pi(224.63~\text{THz})$, $\omega_{ep} = 0.36$~Trad/s, $\alpha_e = 500\times 10^9$, $\omega_{m0} = 2\pi(224.40~\text{THz})$, $\omega_{mp} = 0.29$~Trad/s, $\alpha_m = 100\times 10^9$, and excitation frequency $\omega_0 = 2\pi(230~\text{THz})$. The spectrum is normalized to the non-modulated case with $\Delta_p=0$. }\label{Fig:TDvdFloquet}
\end{center}
\end{figure}

Figure~\ref{Fig:TDvdFloquet} shows several examples of the transmission and reflection spectrum for different modulation depths and frequencies. While the modulation frequency $\omega_p$ only fixes the location of the newly generated harmonic components, the modulation depth $\Delta_p$ controls the relative strengths of the harmonics, compared to the fundamental frequency of excitation. An excellent agreement is observed with FDTD solver, for all the cases considered, validating the proposed procedure. Minor discrepancies are observed between the two, which are attributed to several factors. Firstly, the Floquet solution assumed zero losses here\footnote{only for simplicity, and clearer  compact description of the associated matrices.}, while the FDTD time-domain results consider finite losses. Secondly, the Floquet solution is the steady state response of the metasurface, while the FDTD solver takes into account the dispersion based distortion of the input pulse, which is naturally not taken into account in the Floquet analysis in \eqref{Eq:FieldConstruction}.

\begin{figure}[htbp]
\begin{center}
\psfrag{A}[c][c][0.8]{frequency, $\omega/\omega_0$}
\psfrag{B}[c][c][0.8]{Modulation frequency $\omega_p/\omega_0$}
\psfrag{D}[l][c][0.8]{$20\log|E_t(z= 0_+,\omega)|$}
\psfrag{C}[c][c][0.8]{$20\log|E_t(\omega_0)|$}
\psfrag{E}[r][c][0.8]{$\boxed{\chi_\text{ee} = \chi_\text{mm},~\Delta_p = 0.1}$}
\includegraphics[width=\columnwidth]{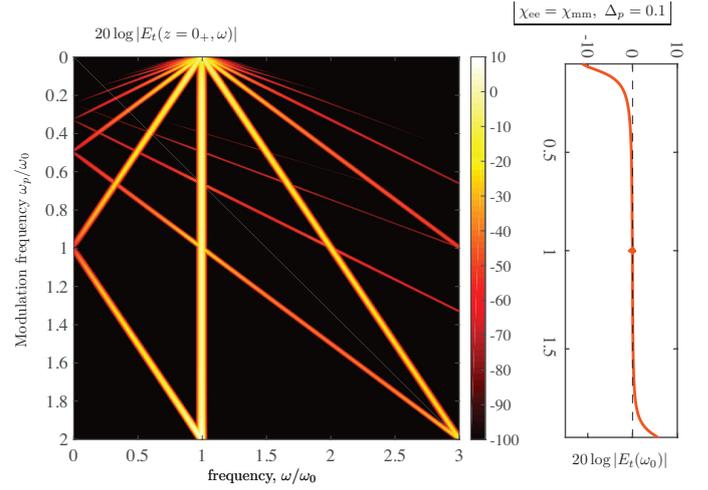}
\caption{Transmission spectrum of a time-only modulated metasurface for a varying modulation frequency and fixed modulation depth, $\Delta_p$. The metasurface is assumed to be matched with $\chi_\text{ee} = \chi_\text{mm}$.}\label{Fig:Scan}
\end{center}
\end{figure}

While Fig.~\ref{Fig:TDvdFloquet} showed results for few discrete points for $\Delta_p$ and $\Delta_m$, Fig.~\ref{Fig:Scan} shows a spectral map for a continuously varied modulation frequency $\omega_p$, for a given modulation depth $\Delta_p$. For simplicity, and considering the practical point of view, a perfectly matched metasurface with $\tilde{\chi}_\text{ee} = \tilde{\chi}_\text{mm}$ is assumed exhibiting zero reflections. The spectral map provides interesting and useful information about the interaction of the metasurface with the input fields. For example, at $\omega_p \approx  2\omega_0$, a substantial amplification of the fundamental frequency $\omega_0$ is observed, which is also the dominant spectral component in the transmitted fields. This is reminiscent of wave amplifications in diverse class of mechanical and electromagnetic parametric systems, where specific parameters of the system are modulated at twice the excitation frequency leading to wave instabilities \cite{TamirST}\cite{OlinerST}.

\begin{figure}[htbp]
\begin{center}
\psfrag{A}[c][c][0.8]{frequency, $\omega/\omega_0$}
\psfrag{B}[c][c][0.8]{$x~\mu$m}
\psfrag{C}[c][c][0.8]{$20\log|E_t(x, \omega)|$}
\psfrag{D}[c][c][0.8]{$\angle E_t(x, \omega)$~rad}
\psfrag{E}[c][c][0.8]{$\boxed{\Delta_p = 0.025$, $f_m = 0.1f_0}$}
\includegraphics[width=\columnwidth]{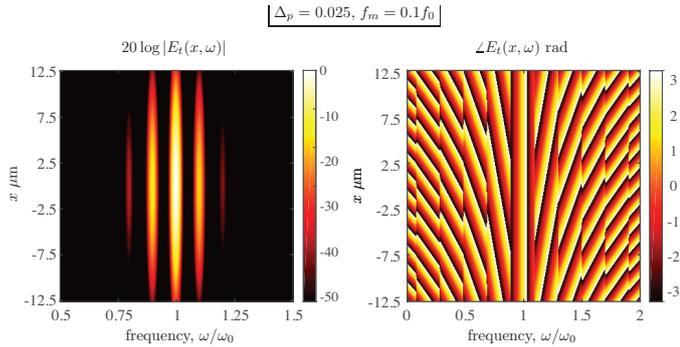}
\caption{Transmission spectrum (amplitude and phase) of a space-time modulated metasurface, when incident with a normally incident Gaussian beam. The beam is given by $E_0(x) = \exp\{-(x/2w_x)^2\}$, with $w_x = 5~\mu$m. The metasurface size $\ell = 25~\mu$m and the excitation frequency $f_0 = 230$~THz. The pumping spatial frequency $\beta_p = 5\pi/\ell$.}\label{Fig:Example}
\end{center}
\end{figure}

Next, an example of a space-time modulated metasurface is considered. Due to the spatial dependence of the susceptibilities following \eqref{Eq:w0Modulation}, the metasurface is spatially discretized and the Floquet analysis is performed at each $x$ location to determine the transmitted and reflected fields, $\tilde{E}_t(x, z=0_+, \omega)$ and  $\tilde{E}_r(x, z=0_+, \omega)$, respectively. An example is shown in Fig.~\ref{Fig:Example}, for an input Gaussian beam, showing the spectrum of the transmitted fields at the output of the metasurface at each location, with the mismatched parameters of Fig.~\ref{Fig:TDvdFloquet}. The Gaussian waveform nature is clearly observed for each spectral component. More interestingly, the phase profile across the metasurface as a function of frequency is also shown in Fig.~\ref{Fig:Example}. A linear phase gradient with different spatial slopes, $d\angle E_t(x)/dx$, across the metasurface is clearly evident for each harmonic component, except the fundamental frequency. Therefore, each of these harmonics are expected to refract at different angles. Similar observations are also made for reflected fields (not shown here).

\begin{figure*}[htbp]
\begin{center}
\psfrag{x}[c][c][0.8]{$x~\mu$m}
\psfrag{z}[c][c][0.8]{$z~\mu$m}
\psfrag{A}[c][c][0.8]{$20\log|E(\omega_0 - \omega_p)|$}
\psfrag{B}[c][c][0.8]{$20\log|E(\omega_0)|$}
\psfrag{C}[c][c][0.8]{$20\log|E(\omega_0 + \omega_p)|$}
\psfrag{k}[c][c][0.8]{spatial frequency $k_x\times 10^{-6}$}
\psfrag{y}[c][c][0.8]{$\mathcal{F}_x\{E(x, z= 0_{\pm}, \omega_0)\}^2$}
\psfrag{n}[c][c][0.8]{$k_{x,\text{in}}$}
\psfrag{m}[c][c][0.8]{$k_{x, -1}^\text{peak}$}
\psfrag{p}[c][c][0.8]{$k_{x, +1}^\text{peak}$}
\psfrag{r}[c][c][0.6]{$\boxed{\theta_{-1} = -8.67^\circ}$}
\psfrag{s}[c][c][0.6]{$\boxed{\theta_{0} = 0^\circ}$}
\psfrag{t}[l][c][0.6]{$\boxed{\theta_{+1} = +7.08^\circ}$}
\includegraphics[width=1.75\columnwidth]{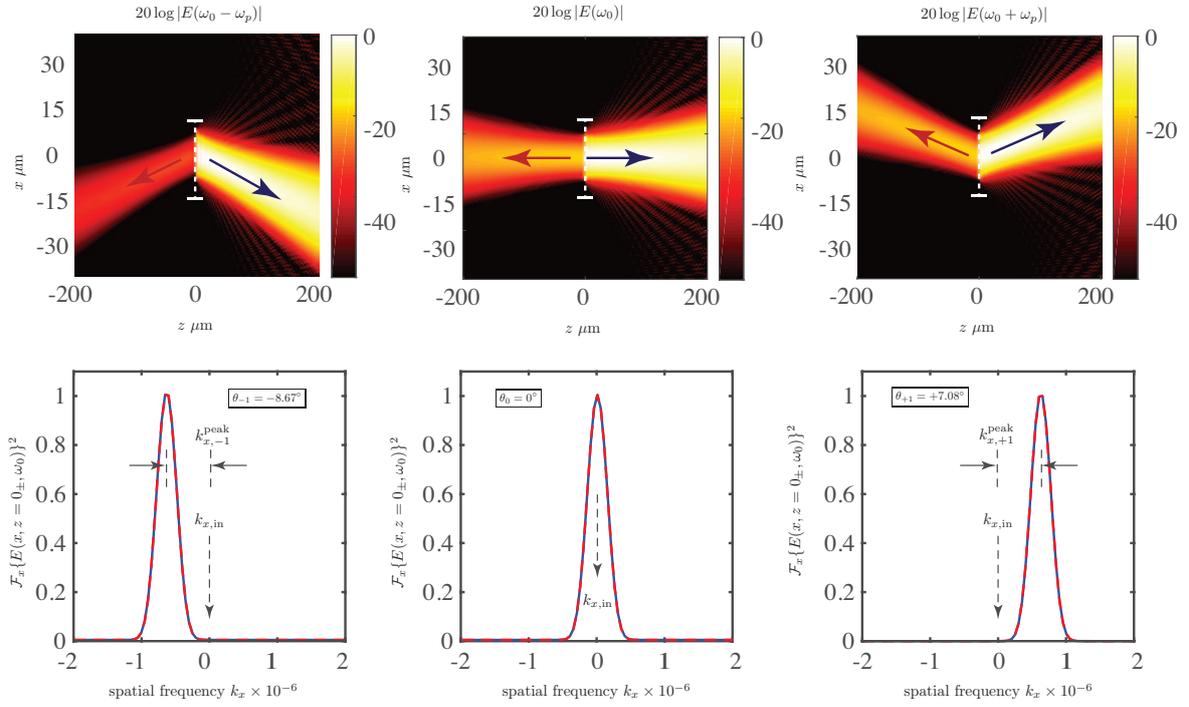}
\caption{Steady-state scattered fields in the transmission and reflection region corresponding to the space-time modulated metasurface of Fig.~\ref{Fig:Example}, obtained using analytical Fourier propagation method, for the first up-converted and down-converted spectral component along with the fundamental, in agreement with the predictions of \eqref{Eq:BetapRefract}. Also shown are the spatial Fourier transforms of the individual harmonics with solid and dashed curves corresponding to transmitted and reflected fields, respectively. All fields are normalized to their own maximum.}\label{Fig:FourierFields}
\end{center}
\end{figure*}

To confirm the refraction angles of the transmitted and reflected fields from the metasurface, the output fields at a specific harmonic frequency $\omega_n$, are then Fourier propagated in free-space as \cite{Goodman_Fourier_Optics}

\begin{subequations}
\begin{equation}
\tilde{E}_t(x, z, \omega_n) = \mathcal{F}_x^{-1}\left[\mathcal{F}_x\{ E_t(x, z_+, \omega_n)\}\exp\{-jk_{z,n}z\}\right]\notag
\end{equation}
\begin{equation}
\tilde{E}_r(x, z, \omega_n) = \mathcal{F}_x^{-1}\left[\mathcal{F}_x\{ E_r(x, z_+, \omega_n)\}\exp\{+jk_{z,n}z\}\right]\notag
\end{equation}
\end{subequations}

\noindent where $k_{z,n} = \sqrt{k_n^2 - k_x^2}$, is the free-space wave number in the $z$ direction. Fig.~\ref{Fig:FourierFields} shows the resulting field in the $x-z$ plane for the first down-converted $(\omega_0-\omega_p)$ and first up-converted harmonic $(\omega_0 + \omega_p)$, in addition to the fundamental frequency $\omega_0$. As expected from Fig.~\ref{Fig:Example} and consistent with \eqref{Eq:BetapRefract}, the fundamental frequency $\omega_0$ propagates without any refraction. On the other hand, the down-converted and up-converted harmonics are refracted in the lower and upper half of the $x-z$ plane, respectively. Their reflection angles using transverse wavenumbers are given by

\begin{equation}
\theta(\omega_n = \omega_0 + n\omega_p) = \sin^{-1}\left[\frac{k_{x, n}^\text{peak}}{k_n}\right].
\end{equation}

\noindent where $k_{x, n}^\text{peak}$ is the peak location of the spatial spectrum of the output fields, $k_n$ is the wavenumber of the $n^\text{th}$ harmonic, and the angle $\theta$ is measured from the normal of the metasurface. For the $n=\pm 1$ harmonics, the refraction angles are found to be $\theta =  8.67^\circ$ and $\theta =  7.08^\circ$, respectively, for both transmission and reflection, with a very good agreement with the theoretical values of $8.33^\circ$ and $6.81^\circ$ obtained from \eqref{Eq:BetapRefract}.

\section{Conclusions}

A rigorous Floquet mode analysis has been proposed for a zero thickness space-time modulated Huygens' metasurface to model and determine the strengths of the harmonics of the scattered fields. The proposed method is based on GSTCs treating the metasurface as a spatial discontinuity. The metasurface has been modelled using space-time varying resonant frequencies of the associated electric and magnetic surface susceptibilities, $\chi_\text{ee}$ and $\chi_\text{mm}$, respectively. The unknown scattered fields have been expressed in terms of Floquet modes, which when used with the GSTCs, led to a system of field matrix equations. The resulting set of linear equations were then solved to determine the transmitted and reflected scattered fields. Using an FDTD solver, the proposed method is validated and confirmed for several examples of modulation depths ($\Delta_p$) and frequencies ($\omega_p$). Finally, the computed steady-state scattered fields are Fourier propagated analytically, for convenient visualization of refracted harmonics. The proposed method is fast, simple and versatile, and is expected to be a useful tool in determining the steady-state scattered fields from a space-time modulated Huygen's metasurface, for arbitrary modulation frequencies and depths.

\section{Appendix}

\subsection*{Summary of Finite-Difference Time Domain (FDTD) Formulation of Space-Time Modulated Metasurface \cite{Smy_Metasurface_Linear}\cite{Stewart_Metasurface_STM}}

Consider a space-time modulated metasurface of Fig.~\ref{Fig:Problem} with the plane-wave input and output fields described in \eqref{Eq:FieldConvention}. The field equations governing the transmitted and reflected fields are the time-domain Lorentz relations \eqref{Eq:TD_Lorentz} and the GSTC equations \eqref{Eq:GSTCEquations}.

Each Lorentz resonator equation of \eqref{Eq:TD_Lorentz} corresponding to incident, transmitted and reflected fields, is a second order differential equation. By introducing an auxiliary variable, each second-order differential equation can be decomposed into two first-order differential equations. For instance, the electric and magnetic polarizabilities corresponding to the incident fields are given by

\begin{subequations}\label{Eq:AuxilliaryLZ}
\begin{equation}
\omega_{e0}\bar{Q}_0 = \frac{dQ_0}{dt} + \alpha_e Q_0, 
\end{equation}
\begin{equation}
\frac{d\bar{Q}_0}{dt}  +  \bar{Q}_0\frac{1}{\omega_{e0}}\frac{d\omega_{e0}}{dt}  + \omega_{e0} Q_0= \frac{ \omega_{ep}^2}{\omega_{e0}} E_0
\end{equation}
\begin{equation}
\omega_{m0}\bar{M}_0 = \frac{dM_0}{dt} + \alpha_m M_0,
\end{equation}
\begin{equation}
\frac{d\bar{M}_0}{dt}  +  \bar{M}_0\frac{1}{\omega_{m0} }\frac{d\omega_{m0}}{dt}  + \omega_{m0} M_0 = -\frac{\omega_{mp}^2}{\eta_0\omega_{m0} } E_0,
\end{equation}
\end{subequations}

\noindent where $\bar{Q}_0$ and $\bar{M}_0$ are the two unknown auxiliary variables in addition to $Q_0$ and $M_0$, for a specified input field $E_0$. Similar set of equations, can be developed for the transmitted and reflected fields in terms of unknown fields $E_t$ and $E_r$. They all represent a set of 12 equations, with 12 auxiliary unknowns and 2 primary unknowns. All the scattered fields and their respective polarizations are coupled to each other through the GSTC equations, which can be written in terms of new auxiliary variables as 

\begin{subequations}\label{GSTCEquationsFDTD}
\begin{equation}
\begin{split}
\omega_{m0}(\bar{M}_0 + \bar{M}_t+ \bar{M}_r) - \alpha_m ({M}_0 + {M}_t+ {M}_r)  \\ = \frac{2}{\mu_0} (E_t - E_r- E_0), 
\end{split}
 \end{equation}
\begin{equation}
\begin{split}
\omega_{e0}(\bar{Q}_0 + \bar{Q}_t+ \bar{Q}_r) - \alpha_e ({Q}_0 + {Q}_t+ {Q}_r)  \\ = \frac{2}{\epsilon_0\eta_0}  (E_0 - E_t - E_r).
\end{split}
 \end{equation}
\end{subequations}

\noindent Equations~\eqref{Eq:AuxilliaryLZ} and \eqref{GSTCEquationsFDTD}, finally represents 14 equations with 14 unknowns, and can be expressed in a matrix form as

\begin{align}\label{Eq:MatrixEquation}
	&\overbrace{\left[
	 \begin{array}{ccc}
	 	\mathbf{0} & \mathbf{0} &  \mathbf{0} \\
		\mathbf{0} & \mathbf{0} &  \mathbf{0}\\
		\mathbf{W_1} & \mathbf{0} &  \mathbf{0} \\
		\mathbf{0} & \mathbf{T_1} &  \mathbf{0} \\
		 \mathbf{0}  & \mathbf{0} &  \mathbf{T_1} \\
	\end{array} 
	\right]}^{\mathbf{[C]}}
	\frac{d[\mathbf{V}]}{dt}\notag\\
	&+ 
	\overbrace{\left[
	 \begin{array}{ccc}
	 	\mathbf{A_1} & \mathbf{A_2} &  \mathbf{A_3} \\
		\mathbf{B_1} & \mathbf{B_2} &  \mathbf{B_3}\\
		\mathbf{W_2}(t) & \mathbf{0} &  \mathbf{0} \\
		\mathbf{0} & \mathbf{T_2}(t) &  \mathbf{0} \\
		 \mathbf{0}  & \mathbf{0} &  \mathbf{T_2}(t) 
	\end{array} 
	\right]}^{\mathbf{[G(t)]}}
	[\mathbf{V}]=\mathbf{[E(t)]},
\end{align}

\noindent where $[\mathbf{V}]$ is the solution vector containing the transmission and reflection field $E_t$ and $E_r$, and $[\mathbf{G(t)}]$ contains the exact description of the metasurface. For a space-time modulated metasurface considered here, the resonant frequencies of the Lorentzian susceptibilities are assumed to be a function of both space and time, i.e. $\omega_{e0}(x,t)$ and $\omega_{m0}(x,t)$. The above matrix equations can be re-written in a complex matrix equation form as

\begin{equation}
\mathbf{[C]}\frac{d\mathbf{[V]}}{dt} + \mathbf{[G(t)]} \mathbf{[V]} = \mathbf{[E(t)]},
\label{Eq:TVMetasurface_Equation}
\end{equation}

\noindent which now can be easily solved using standard finite-difference technique based on trapezoidal integration as

\begin{align}
\mathbf{[V]}_i    =  \left(\mathbf{[C]} +  \frac{\Delta t}{2}\mathbf{[G]}_i \right )^{-1} &\left[\Delta t\frac{ \mathbf{[E]_i} +  \mathbf{[E]_{i-1}}}{2} \right. \notag\\
& \left. + \left(\mathbf{[C]} - \frac{\Delta t}{2} \mathbf{[G]}_i\right) \mathbf{[V]}_{i-1}\right].\notag
\end{align}

\bibliographystyle{IEEETran}
\bibliography{Gupta_Analytical_STModulated_TAP_2017}

\end{document}